\documentclass[showpacs,aps,prl,twocolumn,superscriptaddress,preprintnumbers,nofootinbib]{revtex4-1}
\usepackage{amsmath}
\usepackage{epsfig}
\usepackage{subfigure}
\usepackage{float}
\usepackage{verbatim}
\usepackage{axodraw4j}
\usepackage{pstricks}
\usepackage{color}
\usepackage{slashed}
\usepackage{multirow}
\usepackage{pstricks}
\usepackage{color}
\usepackage{booktabs}
\usepackage{subfigure}
\usepackage{epstopdf}
\usepackage{dsfont}

\newcommand{\lag}{{\cal L}}

\renewcommand{\section}[1]{{{\textit{ #1.}}---}}

\begin{document}


\title{Single Top Production at Next-to-Leading Order in the Standard Model
Effective Field Theory}

\author{Cen Zhang}
\affiliation{
Department of Physics, Brookhaven National Laboratory, Upton, New York, 11973, USA
}

\begin{abstract}
Single top production processes at hadron colliders provide information on the
relation between the top quark and the electroweak sector of the standard
model.  We compute the next-to-leading order QCD corrections to the
three main production channels: $t$-channel, $s$-channel and $tW$
associated production, in the standard model including operators up to
dimension-six. The calculation can be matched to parton shower programs and can
therefore be directly used in experimental analyses. The QCD corrections are
found to significantly impact the extraction of the current limits on the
operators, because both of an improved accuracy and a better precision of the
theoretical predictions. In addition, the distributions of some of the key
discriminating observables are modified in a nontrivial way, which could change
the interpretation of measurements in terms of UV complete models.  

\end{abstract}


\maketitle

\section{Introduction}At high-energy colliders, physics beyond the standard
model (SM) is searched for either by looking for evidence of new particles or
for deviations in the predicted interactions between the SM particles. In the
latter effort the top quark plays a special role: thanks to its large mass it
can naturally probe high scales and in particular the electroweak symmetry
breaking sector.  A general theoretical framework where the experimental
information on the interactions and possible deviations can be consistently and
systematically interpreted is provided by the SM effective field theory (SMEFT)
approach \cite{Weinberg:1978kz,Buchmuller:1985jz,Leung:1984ni}. The SMEFT
Lagrangian corresponds to that of the SM augmented by  higher-dimensional
operators that respect the symmetries of the SM.  It provides a powerful
approach to identify observables where deviations could be expected in the top
sector \cite{Cao:2007ea,Zhang:2010dr,Degrande:2010kt}.  Besides and more
importantly, it allows a global interpretation of measurements coming from
different processes and experiments \cite{Zhang:2012cd,Durieux:2014xla,
Buckley:2015nca,Buckley:2015lku}, which can be consistently evolved up to new
physics scales, and provide hints to specific models at high scales.

Given the results of the LHC run I \cite{Agashe:2014kda}, expectations  from
run II on the attainable precision of the top-quark couplings are very high.
Theoretical predictions that are at least as accurate and precise as the
experimental projections are thus required. This motivates the calculation  of
higher-order corrections. In this work, we focus on the single-top production
processes.  At the LHC, single-top production proceeds through three main
channels: $t$-channel, $s$-channel and $tW$ associated production.  They are
ideal for probing the top-quark couplings to the electroweak sector of the SM,
and can provide key and complementary information to that coming from top-quark
decay.  To this aim we promote the single-top predictions, for the first time,
to next-to-leading order (NLO) in QCD in the SMEFT, and study their impact on
the interpretation of measurements. 

The main results of this work can be summarized as follows.  First, we show
that QCD corrections not only affect total cross sections and reduce their
uncertainties, but also impact the distributions of key observables, in such a
way that the interpretation of possible deviations from the SM would lead to
quite different UV complete models.  Moreover, these corrections cannot be
captured by either the $K$-factors or the renormalization group (RG)
improvements of the Wilson coefficients.  Second, we demonstrate that a new
type of scale uncertainty in EFT, coming from the running and mixing of
dimension-six terms, needs to be considered and can be reduced by including QCD
corrections.  Finally, by matching our NLO computation to a parton shower (PS)
program, predictions can be obtained through an event generator that can be
used directly in experimental simulations, to design optimized analyses that
can maximize the sensitivity to new physics.  

\section{Effective operators}
\begin{figure}[htb]
	\begin{center}
		\includegraphics[width=.88\linewidth]{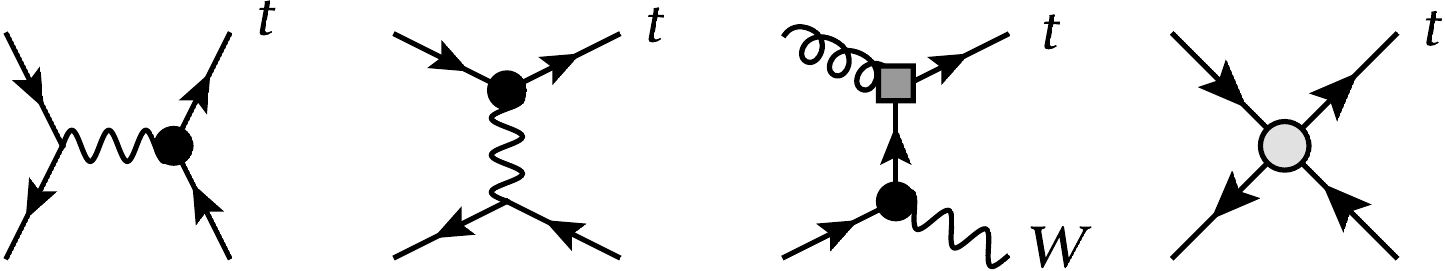}
	\end{center}
	\caption{Representative leading order (LO) diagrams for all three
		single-top channels.  Vertices with a black dot can be modified
		by $O_{\phi Q}^{(3)}$ and $O_{tW}$, while that with a square is
		modified by $O_{tG}$. The last diagram
	comes from $O_{qQ}^{(3)}$.}
		\label{fig:fdiagram}
\end{figure}

In the EFT approach deviations from the SM are
captured by effective operators.  Up to dimension six, four operators are
relevant
\cite{Cao:2007ea,AguilarSaavedra:2008zc,Zhang:2010dr}:
\newcommand{\FDFI}{\left(\varphi^\dagger\overleftrightarrow{D}^I_\mu\varphi\right)}
\begin{flalign}
  &O_{\varphi Q}^{(3)}
  =i\frac{1}{2}y_t^2 \FDFI (\bar{Q}\gamma^\mu\tau^I Q)
  \label{eq:Ofq3}
  \\
  &O_{tW}=y_tg_W(\bar{Q}\sigma^{\mu\nu}\tau^It)\tilde{\varphi}W_{\mu\nu}^I
  \\
  &O_{tG}=y_tg_s(\bar{Q}\sigma^{\mu\nu}T^At)\tilde{\varphi}G_{\mu\nu}^A
  \\
  &O_{qQ,rs}^{(3)}=(\bar q_r\gamma_\mu\tau^Iq_s)(\bar Q\gamma^\mu\tau^IQ)
  \label{eq:Otf}
\end{flalign}
Here $q_r$ and $q_s$ are the quark doublet fields in the first two
generations, while $Q$ is in the third generation. $r,s$ are flavor indices.
$\varphi$ is the Higgs doublet.  $g_W$, $g_Y$ and $g_s$ are the SM gauge coupling
constants.  $y_t$ is the top-quark Yukawa coupling, defined by its pole mass.
The effective Lagrangian is
\begin{equation}
	\lag_\mathrm{eff}=\lag_\mathrm{SM}+\sum_i
	\frac{C_{i}}{\Lambda^2}O_{i}+H.c. \,,
\end{equation}
where $\Lambda$ is the expected scale of new physics. $C_i$ is the
coefficient to parametrize the deviation from $O_i$.
In this work we assume flavor universality in the first two generations,
defining $O_{qQ}^{(3)}=O_{qQ,11}^{(3)}+O_{qQ,22}^{(3)}$.
Dimension-six operators affect all three channels.  Corresponding diagrams are
shown in Fig.~\ref{fig:fdiagram}.

The operators $O_{tG}$ and $O_{tW}$ have nonzero anomalous dimensions at
$\mathcal{O}(\alpha_s)$, given by
\cite{Jenkins:2013zja,Jenkins:2013wua,Alonso:2013hga,Zhang:2014rja}
\begin{flalign}
	\frac{dC_i}{d\log\mu}=\frac{\alpha_s}{\pi}\gamma_{ij}C_j\ ,
	\quad
	\gamma= \frac{1}{3}\left(
	\begin{array}{cc}
		1 & 0 \\
		2 & 2
	\end{array}
	\right)
	\label{eq:adm}
\end{flalign}
This matrix controls the running and mixing of the operators and can be used to
evolve them from scale $\Lambda$ down to the scales of the
measurements.  

\section{Calculation}\label{sec:imp}\label{sec:implementation}The NLO
automation is implemented and validated in the {\sc MadGraph5$\_$aMC@NLO}
framework \cite{Alwall:2014hca}, with the help of a series of packages,
including {\sc FeynRules} and {\sc NLOCT} \cite{Alloul:2013bka, Degrande:2014vpa,
Degrande:2011ua,deAquino:2011ub,Hirschi:2011pa,Frederix:2009yq,Frixione:2002ik}.
A model in the Universal {\sc FeynRules} Output format \cite{Degrande:2011ua}
is built at NLO, allowing for simulating a variety of processes
important for top-coupling measurements.  In this work we only focus on
single-top processes, but other promising (and more complicated) channels, such
as $t\bar tZ/W/\gamma$ and $tjZ/\gamma$, are all made available at NLO in EFT
with PS.  In Ref.~\cite{Bylund:2016phk} we have discussed the physical results
for $t\bar tZ/W/\gamma$ processes.  More details of this implementation will be
presented in a separate work~\cite{CEN}.

We adopt $\overline{MS}$ with five-flavor running in $\alpha_s$ with the
top-quark subtracted at zero momentum transfer \cite{Collins:1978wz}.
Additional contributions to top-quark and gluon-field renormalizations and
$\alpha_s$ renormalization from $O_{tG}$ are included \cite{Franzosi:2015osa}.
For operator coefficients we use $\overline{MS}$ subtraction, with
\begin{align}
	C_{i}^0&\to Z_{ij}C_{j}(\mu')
	\nonumber\\&=
	\left[\mathds{1}+\frac{\alpha_s}{2\pi}
\Gamma(1+\varepsilon)\left( \frac{4\pi\mu^2}{\mu'^2}\right)^\varepsilon
	\frac{1}{\varepsilon_{UV}}\gamma
	\right]_{ij}C_j(\mu')
	\label{eq:zc}
\end{align}
where the anomalous dimension matrix $\gamma$ is given in Eq.~(\ref{eq:adm}).
UV counterterms needed in this work are computed using the above information.
Note that with Eq.~(\ref{eq:zc}) the operators will run with $\mu'$ separately from
the running of $\alpha_s$.  This allows for the dynamical renormalization scale
to be adopted without having to run the operator coefficients.

Results are presented in terms of operators defined at $\mu'=m_t$, i.e.~the log
terms from high scale, $\log\left( \Lambda/m_t \right)$, are already resummed
by evolving operators down to this scale using Eq.~(\ref{eq:adm}).  Thus the
NLO corrections presented here do not include any of such large log terms, and
{\it cannot} be captured by the RG equations. 

\section{Total cross sections}Cross sections, obtained at LO and NLO, can be
parametrized as
\begin{equation}
	\sigma=\sigma_\mathrm{SM}+\sum_i\frac{1\ \mathrm{TeV}^2}{\Lambda^2}
	C_i\sigma^{(1)}_i +\sum_{i\leq j}
	\frac{1\ \mathrm{TeV}^4}{\Lambda^4}C_iC_j\sigma^{(2)}_{ij}+\dots
	\nonumber
	\label{eq:xsec}
\end{equation}
We work up to order $1/\Lambda^2$, and present results for
$\sigma_i^{(1)}$, the interference between an operator $O_i$ and the
SM.   We use NNPDF2.3 parton distributions \cite{Ball:2013hta}. Input parameters are
\begin{align}
	&m_t=172.5\ \mathrm{GeV} & m_Z=91.1876\ \mathrm{GeV}\\
	&\alpha(m_Z)=1/127.9 &G_F=1.16637\times10^{-5} \mathrm{GeV}^{-2}
	\label{eq:input}
\end{align}
Central renormalization and factorization scales are fixed at
$\mu_R=\mu_F=m_t$.  To estimate theoretical uncertainties due to missing
higher orders we perform variations with nine combinations of $(\mu_R,\mu_F)$,
where $\mu_{R,F}$ can take values $m_t/2$, $m_t$ and $2m_t$. 

\begin{figure}[htb]
	\centering
	\begin{center}
		\includegraphics[width=.83\linewidth]{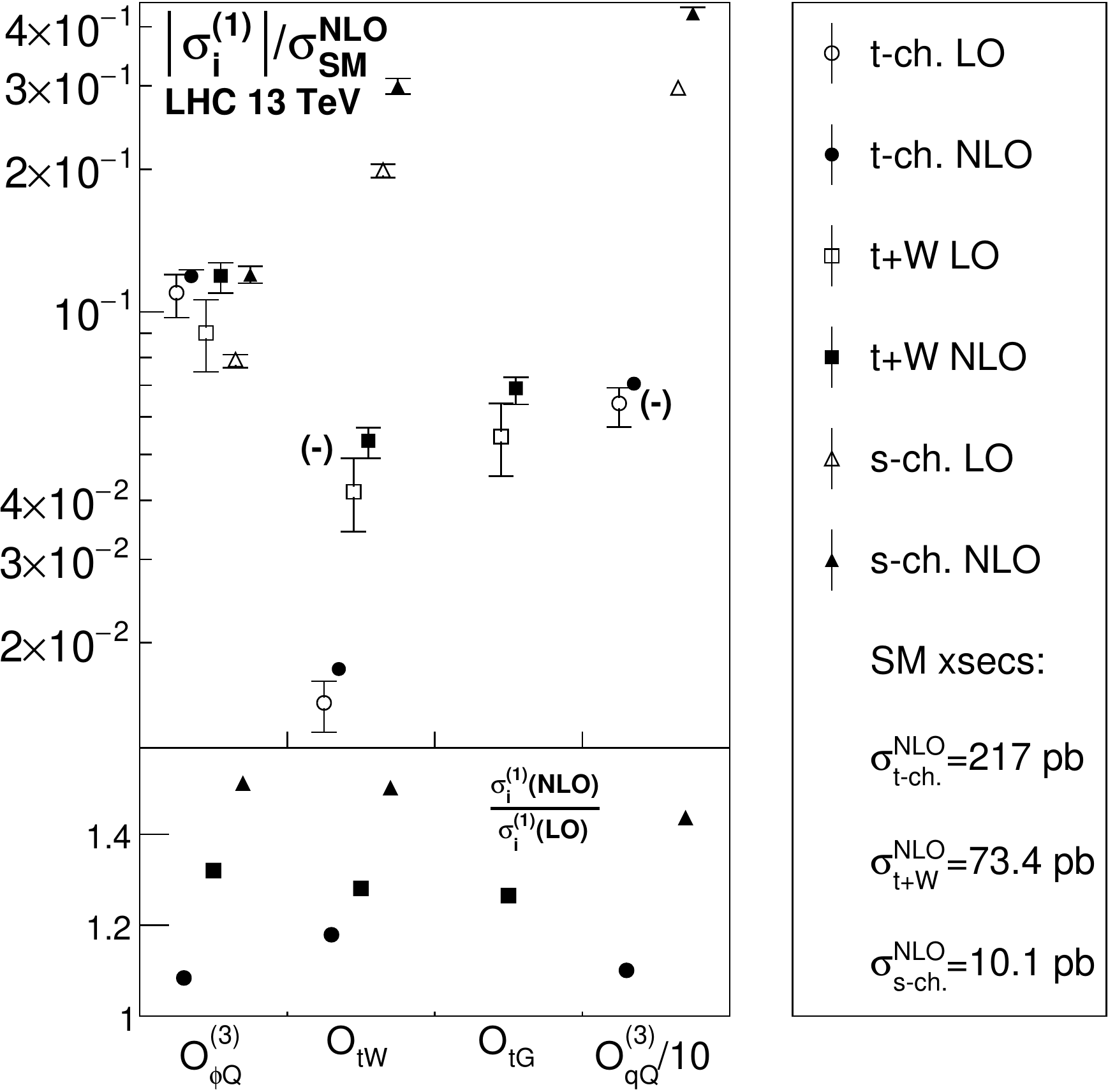}
	\end{center}
	\caption{$r_i=\left|\sigma^{(1)}_i\right|/\sigma_{\mathrm{SM}}^{\mathrm{NLO}}$
	for the three single-top channels.  Both LO and NLO results are shown.
	Error bars indicate scale uncertainties.  $K$-factors are given in the
	lower panel.  Negative contributions are labeled with ``(-)''.
	\label{fig:xsec}}
\end{figure}

Total cross sections (including top and antitop) at LHC 13 TeV are presented
in Fig.~\ref{fig:xsec}.  We plot the ratio between the interference cross
section, $\sigma^{(1)}_i$, and SM NLO cross section,
$r_i=\left|\sigma^{(1)}_i\right|/\sigma_{\mathrm{SM}}^{\mathrm{NLO}}$, for
individual operators $O_i$, in all three channels.  The ratio $r_i$ illustrates
how sensitive a process is to a certain operator, and can be interpreted as the
signal over background ratio.  In the plot, scale uncertainties from the
numerator are given, and in the lower panel we show the $K$-factor of each
operator contribution.  Improved accuracy is reflected by the $K$-factors,
typically ranging from $\sim10\%$ to $\sim50\%$, and improved precision is
reflected by the significantly reduced scale uncertainties.  Furthermore, most
NLO results are outside of the uncertainty range of corresponding LO results,
indicating that QCD corrections are essential for a correct interpretation of
measurements in terms of operators.  For comparison, at 8 TeV the $t$-channel
has been measured at better than $\sim10\%$ level
\cite{Khachatryan:2014iya,ATLAS:2014dja}, and the $t+W$ channel is at about
$20\%$ \cite{CMS:2014efa}.  At the high-luminosity LHC the $t$-channel can
reach $\sim4\%$ \cite{Schoenrock:2013jka}, while the $s$-channel may reach
$\sim15\%$ \cite{Selvaggi:2015sdf}.  NNLO approximate QCD corrections are
available for the SM predictions, and corresponding theoretical uncertainties
are at the percentage level \cite{Kidonakis:2012rm}.

\begin{figure}[htb]
	\centering
	\begin{center}
		\includegraphics[width=.53\linewidth]{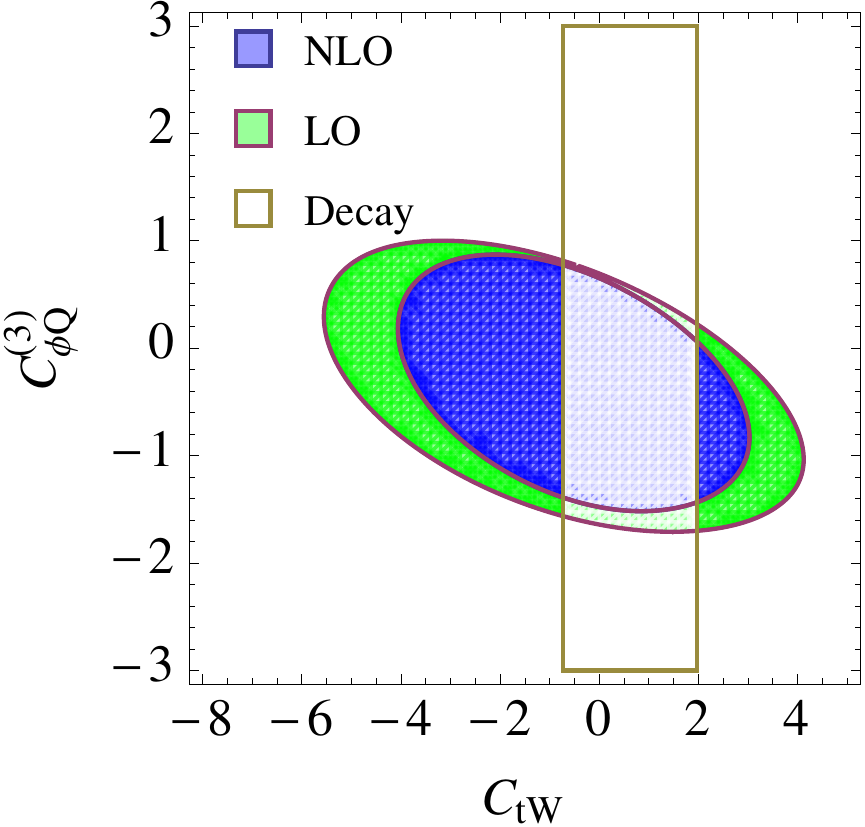}
		\raisebox{-1ex}{\includegraphics[width=.455\linewidth]{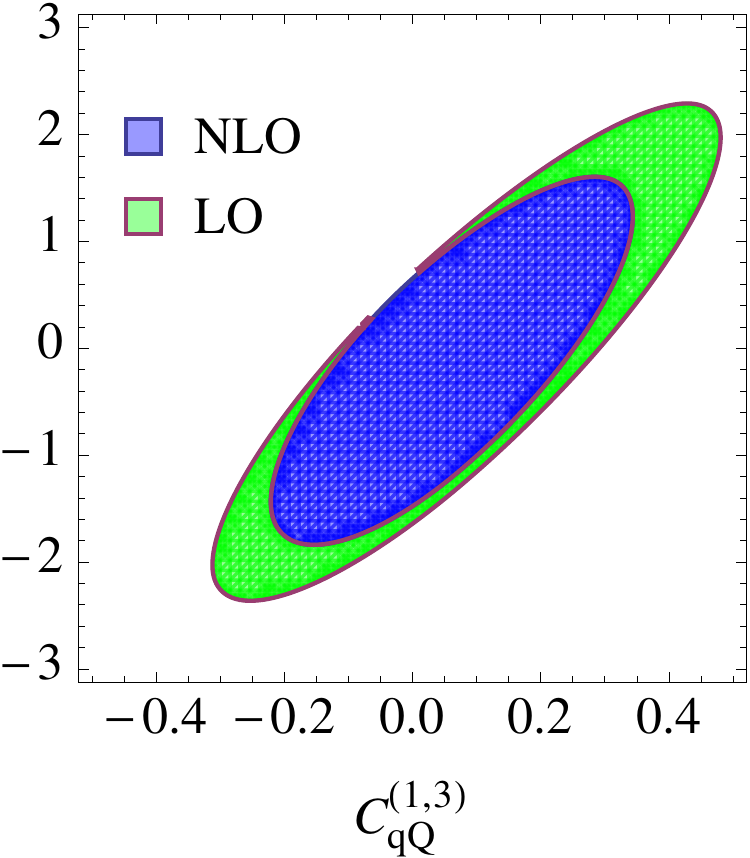}}
	\end{center}
	\caption{95\% limit from single-top measurements, with LO or NLO predictions
		for EFT. Left: $(O_{\phi Q}^{(3)}, O_{tW})$; right: 
		$(O_{\phi Q}^{(3)}, O_{qQ}^{(3)})$.
		Limits from top decay measurements are compared.
		\label{fig:xsecfit}}
\end{figure}

NLO corrections already affect current bounds on the coefficients of the
dimension-six operators.  For illustration we perform two-operator fits, for
$(O_{\phi Q}^{(3)}, O_{tW})$ and for $(O_{\phi Q}^{(3)}, O_{qQ}^{(3)})$, using
cross sections available at the LHC at 8 TeV
\cite{Khachatryan:2014iya, ATLAS:2014dja,CMS:2014efa,Aad:2014aia} with the
state-of-the-art SM prediction \cite{Kidonakis:2012rm} and NLO EFT predictions
from this work.  Limits are improved thanks to better accuracy and precision,
and can be clearly seen in Fig.~\ref{fig:xsecfit}.  For comparison we also show
current limits on $O_{tW}$ from decay measurements
\cite{CMS:2015fja,Fischer:1998gsa,Zhang:2014rja}. (See also
Ref.~\cite{deBlas:2015aea} for RG-induced bounds on top-quark operators.)

\section{Distributions}The QCD corrections have more crucial effects on the
shapes of observables that can be used to identify deviations.  Some key
observables have very distinct distributions that depend on the relative
contribution from different operators.  If any deviation in total cross section
is observed, these observables will determine which operator is the source of
the deviation.  Even without any deviation, including these observables in a
global analysis can help to constrain flat directions.

\begin{figure}[htb]
	\begin{center}
		\includegraphics[width=.89\linewidth]{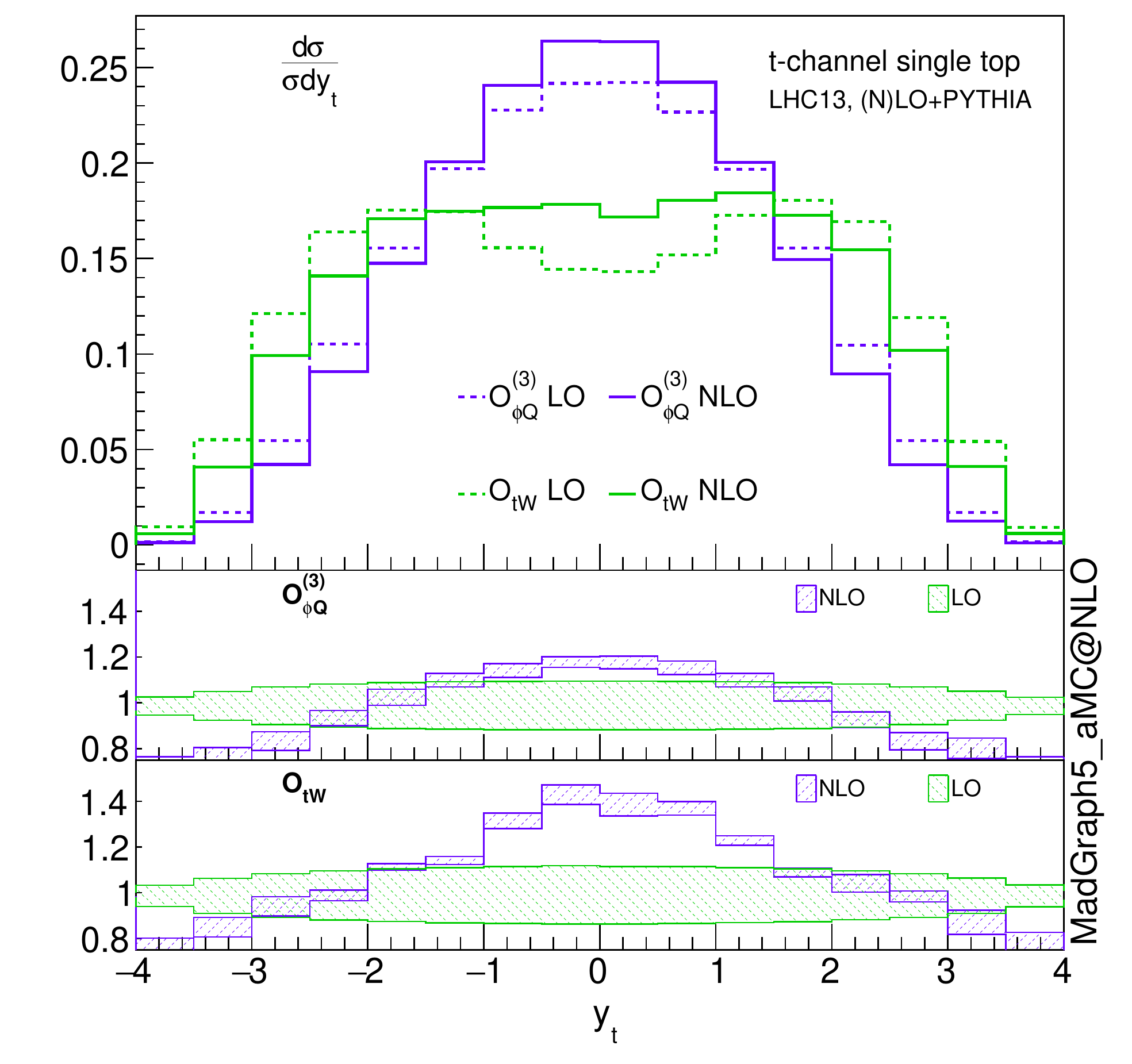}
	\end{center}
	\caption{Normalized rapidity distributions of
	the top quark in $t$-channel single-top production, from $O_{tW}$ and
	$O_{\phi Q}^{(3)}$.  Only the interference with the SM is included.
	Lower panel shows the $K$-factors of individual operators, with scale
	uncertainties.}
	\label{fig:tch:yt}
\end{figure}

In our approach, distributions can be obtained at NLO in QCD with PS simulation
\cite{Frixione:2002ik,Sjostrand:2006za}, and with top quarks decayed keeping
spin correlations \cite{Artoisenet:2012st}.  In Fig.~\ref{fig:tch:yt} we show
the normalized distributions of the top-quark rapidity, $y_t$, in $t$-channel
single-top production, which is an efficient discriminating observable, and has
been measured already \cite{Aad:2014fwa,CMS:2014ika}.  We can see that its
distribution is more forward for $O_{tW}$ while rather central for $O_{\phi
Q}^{(3)}$.  The difference arises already at the parton level due to the
Lorentz structure of $O_{tW}$ suppressing the forward scattering amplitude
\cite{Zhang:2010dr}, and it is diluted at NLO due to real corrections.

Fig.~\ref{fig:tch:yt} also explains why NLO corrections are important when shape
information is used.  It makes both distributions more central, and
missing this correction would lead to an underestimate of the size of the
$O_{tW}$ contribution on one hand and a corresponding overestimate of
$O_{\phi Q}^{(3)}$ on the other.  We find that other variables, including $p_T$
and rapidity of the first non-$b$ jet and of the first $b$ jet, are affected in
a similar way.  Moreover, the theory uncertainty in shapes due to missing QCD
is not captured by varying $\mu_R$ and $\mu_F$.  We thus conclude that
NLO QCD corrections can lead to bias in an EFT analysis, by shifting
the theoretical predictions for the shapes of discriminating observables.

\begin{figure}[htb]
	\begin{center}
		\includegraphics[width=.81\linewidth]{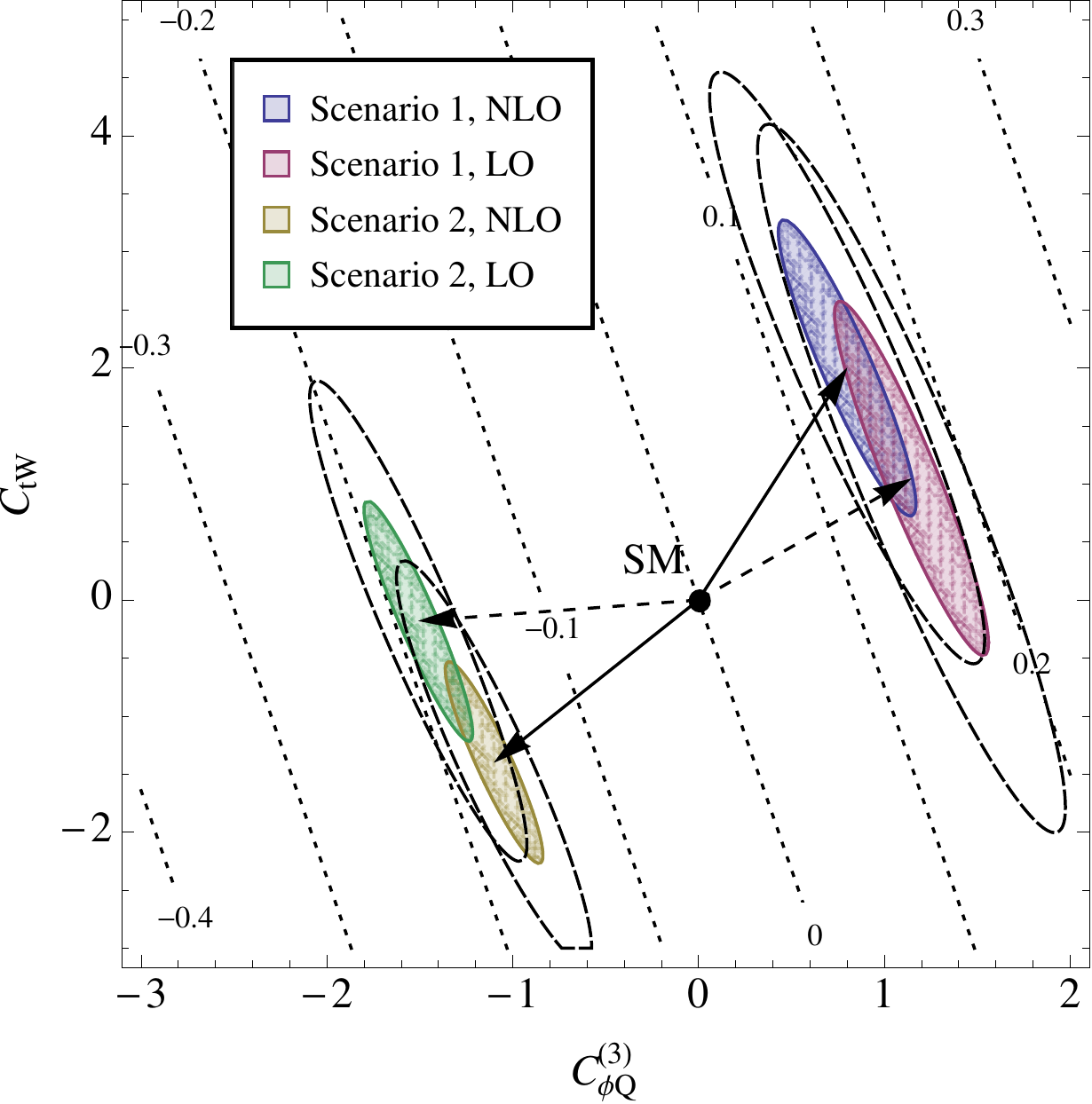}
	\end{center}
	\caption{Two-operator fit using pseudomeasurements on shapes, at 68\%
		confidence level, assuming 5\% uncertainty in each bin.  Dashed
		lines correspond to twice this uncertainty, while dotted
		contours are the relative deviation in total cross section.}
		\label{fig:shapefit}
\end{figure}

To quantify this effect, we consider two benchmark points, (1):
$C_{\phi Q}^{(3)}=0.8$, $C_{tW}=2$, and (2): $C_{\phi Q}^{(3)}=-1.1$,
$C_{tW}=-1.4$, each corresponding to about a 15\% deviation in the total cross
section.  We compute at NLO the distributions of two observables,
$y_t$ and $p_T$ of the first non-$b$ jet, and use the results as pseudodata,
which we consider in 5 bins for $p_{T,j}$ from 20 to 180 GeV and 6 bins for
$|y_t|$ from 0 to 3. We then perform $\chi^2$ fits with LO and NLO predictions
respectively and compare.  Results depend on the combined uncertainty of
experiment and theory.  Current data at LHC 8 TeV correspond to
$\sim10\%$ uncertainty in each bin \cite{CMS:2014ika}.  Foreseeing future
improvements in the analyses, we assume $\sim5\%$ uncertainty in each bin and
we find that the operator coefficients extracted from the fit are
shifted by NLO effects.  This is shown in Fig.~\ref{fig:shapefit}.

The dotted contours in Fig.~\ref{fig:shapefit} represent a constant deviation
in the cross section.  Cross section measurements constrain the direction
orthogonal to these contours.  On the other hand, including shape
information constrains the direction along these lines.
The bias induced by QCD corrections is reflected by the dashed and the
solid arrows, which represent the resulting deviations from the fit, at LO and
NLO respectively.  For example, in the second scenario the central values of
coefficients extracted at LO are $(-1.5,-0.18)$, and become $(-1.1,-1.4)$ at
NLO, and the one-sigma regions have almost no overlap.  This shift is not in the
radial direction corresponding to an overall rescale by the NLO $K$-factor.
Rather, it leads to a different direction of deviation in the $C_{\phi Q}^{(3)}
- C_{tW}$ plane, as clearly indicated by the angle between the dashed and the
solid arrows.

At this point it is important to note that the two operators, $O_{\phi
Q}^{(3)}$ and $O_{tW}$, correspond to different types of new physics
\cite{Cao:2015doa}.  The first operator is likely to be generated by mixing SM
particles with heavy objects such as $W'$ \cite{Hsieh:2010zr,Cao:2012ng} and
heavy quarks \cite{Cacciapaglia:2010vn,Aguilar-Saavedra:2013qpa}; the second
one is loop induced, and typical scenarios include two-Higgs-doublet models
\cite{Grzadkowski:1991nj} and supersymmetric models
\cite{Dabelstein:1995jt,Cao:2003yk,Li:1992ga}.  It follows that a missing
QCD correction will lead us to an incorrect conclusion about the type of UV
physics.

To sum up, there are two kinds of QCD NLO effects for single-top processes.
The first is on total cross sections.  It can be captured by applying a
$K$-factor to LO results, and only affects the magnitude of deviation from the
SM.  The second is on the shapes of discriminator observables.  It {\it cannot}
be captured by a simple $K$-factor, and it affects the {\it direction} in which new
physics deviates from the SM.  Hence it is important because if deviations are
observed in the single-top channel, missing such corrections would lead us to
misinterpret measurements of possible deviations and misconclude the nature of
UV physics.

\section{EFT scale uncertainties}Perturbative calculations performed in SMEFT
suffer from a new source of scale uncertainty: the running and mixing of
operator coefficients.  In our calculation operators are defined at a scale
$\mu'$, separately from $\mu_{R,F}$. This allows us to study this uncertainty
alone, independent of the usual renomalization and factorization scale
uncertainties.

This uncertainty can be estimated with
$\sigma_i^{(1)}(\mu',\mu'_0)\equiv\Gamma(\mu',\mu'_0)_{ji}
\sigma_j^{(1)}(\mu')$; i.e.,~the operator contributions at $\mu'$
evolved back to central scale $\mu'_0$. Here $\Gamma_{ij}$ is the solution to
the RG equations:
\begin{flalign}
	\Gamma_{ij}(\mu',\mu'_0)
	=&\exp\left(
	\frac{-2}{\beta_0}\log\frac{\alpha_s(\mu')}{\alpha_s(\mu'_0)}
	\gamma_{ij}\right)\,,
\end{flalign}
with $\beta_0=11-2/3n_f$, and
$n_f=5$ is the number of running flavors.

\begin{figure}[htb]
	\begin{center}
		\includegraphics[width=.82\linewidth,height=.55\linewidth]{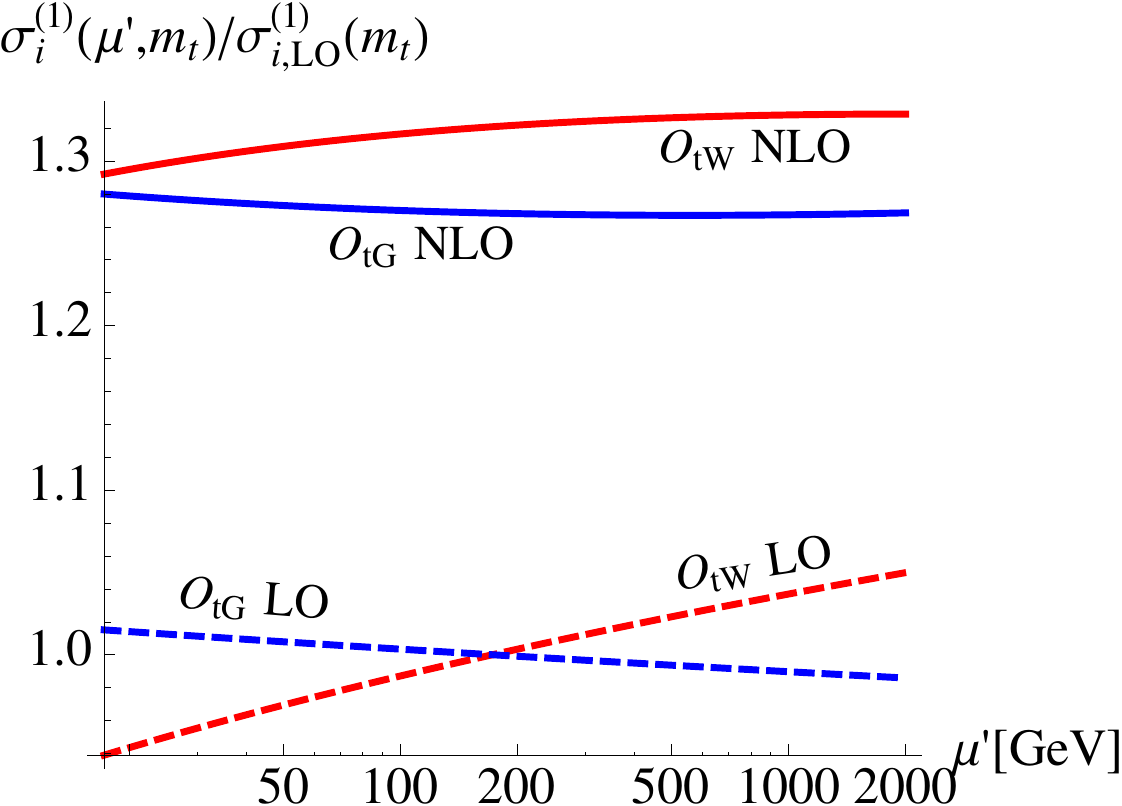}
	\end{center}
	\caption{The $\mu'$ dependence of $\sigma^{(1)}_{tW}(\mu',m_t)$
	and $\sigma^{(1)}_{tG}(\mu',m_t)$ in $tW$ production, normalized
	with $\sigma^{(1)}_{tW,tG}(\mu'=m_t)$ at LO.  The dashed (solid) lines
correspond to LO (NLO) calculation with one-loop running and mixing.}
	\label{fig:twrg}
\end{figure}

For illustration, we present the scale variation in the $tW$ associated
channel.  This process involves both $O_{tW}$ and $O_{tG}$ already at the tree
level, so both the running and the mixing effects are observable.  In
Fig.~\ref{fig:twrg} we show the $\mu'$ dependence of the dimension-six
contribution from $O_{tW}$ and $O_{tG}$, where we choose $\mu'_0=m_t$ as the
central scale, and vary $\mu'$ from $m_t/10$ to 2 TeV, fixing $\mu_R$ and
$\mu_F$.  It is clear from the plot that this kind of scale dependence can be
reduced at NLO, indicating that the leading QCD log terms from the running and
mixing of operator coefficients are cancelled by NLO corrections. We should
point out that there are cases where mixing effects are much more important
than the presented example in Fig.~\ref{fig:twrg}
\cite{
Degrande:2012gr,Grojean:2013kd,Chen:2013kfa,Alonso:2013hga,Elias-Miro:2013eta,
Jung:2014kxa,Englert:2014cva,Zhang:2014ona,deBlas:2015aea,Wells:2015cre}, but
the latter is a proof of principle that the related EFT scale
uncertainties can be taken under control by including the full NLO corrections.

Finally, it is worth pointing out that the RG equations for operators cannot
capture the dominant NLO corrections.  From the plot we can see that the RG
correction to $O_{tW}$ from high scale $\Lambda$ down is negative, while the
complete NLO correction gives a sizeable increase. A reliable result can only
be obtained by carrying out the complete NLO computation.  A similar
observation in the context of Higgs physics has been pointed out by the authors of
Ref.~\cite{Hartmann:2015oia,Hartmann:2015aia}.

\section{Summary}We have presented predictions for single-top processes at NLO
with PS in SMEFT.  Bounds on higher-dimensional
operators are improved thanks to better accuracy and precision.  More
importantly, QCD corrections lead to nontrivial modifications to the shapes of
the most powerful discriminating observables.  If new physics shows up in
single-top processes, missing such corrections would change the interpretation
of the measurements and lead us to bias our interpretations in terms of new
physics models.  We have also demonstrated that the scale uncertainties
associated with the running and mixing of operator coefficients should be
considered, and can be reduced by including NLO corrections.  

Our results should be used in experimental simulations, as they are important
for interpreting measurements, and are available as an NLO+PS event generator.
With more accurate and precise EFT simulation and uncertainties under control,
SM deviations can now be analyzed in a top-down way, designing new analyses to
maximize sensitivity and allowing for a more efficient approach to the study of
the top-quark interactions.

\bigskip
I would like to thank F.~Maltoni for constantly supporting projects on EFT at
NLO in QCD.  I am grateful for valuable discussions with S.~Dawson and
M.~Selvaggi.  I would like to thank E.~Vryonidou for pointing out typos.
This work is supported by U.S.~Department of Energy under Grant
No.~DE-SC0012704.
\bibliography{bib}
\bibliographystyle{apsrev4-1_title}

\end{document}